# Constrained Automated Mechanism Design for Infinite Games of Incomplete Information

**Yevgeniy Vorobeychik**
University of Michigan
Computer Science & Engineering
Ann Arbor, MI 48109-2121 USA
yvorobey@umich.edu

**Daniel M. Reeves**
Yahoo! Research
45 W 18th St, 6th Floor
New York, NY 10011-4609 USA
dreeves@yahoo-inc.com

**Michael P. Wellman**
University of Michigan
Computer Science & Engineering
Ann Arbor, MI 48109-2121 USA
wellman@umich.edu

**Abstract**

We present a functional framework for automated mechanism design based on a two-stage game model of strategic interaction between the designer and the mechanism participants, and apply it to several classes of two-player infinite games of incomplete information. At the core of our framework is a black-box optimization algorithm which guides the selection process of candidate mechanisms. Our approach yields optimal or nearly optimal mechanisms in several application domains using various objective functions. By comparing our results with known optimal mechanisms, and in some cases improving on the best known mechanisms, we provide evidence that ours is a promising approach to parametric design of indirect mechanisms.

## 1 Motivation

While the field of Mechanism Design has been quite successful within a wide range of academic disciplines, much of its progress came as a series of arduous theoretical efforts. In its practical applications, however, successes have often been preceded by a series of setbacks, with the drama of auctioning radio spectrum licenses that unfolded in several countries providing a powerful example [McMillan, 1994].

A difficulty in practical mechanism design that has been especially emphasized is the unique nature of most practical design problems. Often, this uniqueness is manifest in the idiosyncratic nature of objectives and constraints. For example, when the US government tried to set up a mechanism to sell the radio spectrum licenses, it identified among its objectives promotion of rapid deployment of new technologies. Additionally, it imposed a number of ad hoc constraints, such as ensuring that some licenses go to minority-owned and women-owned companies [McMillan, 1994].

Thus, a prime motivation for Conitzer and Sandholm's automated mechanism design work [Conitzer and Sandholm, 2002, 2003a] was to produce a framework for solving mechanism design problems computationally given arbitrary objectives and constraints. We likewise pursue this goal, and seek in addition to avoid reliance on direct truthful mechanisms. This reliance has at its core the Revelation Principle [Myerson, 1981], which states that the outcome of any given mechanism can still be achieved if we restrict the design space to mechanisms that induce truthful revelation of agent preferences. While theoretically sound, there have been criticisms of the principle on computational grounds, for example, those leveled by Conitzer and Sandholm [2003b]. It is also well recognized that if the design space is restricted in arbitrary ways, the revelation principle need not hold. While the computational criticisms can often be addressed to some degree within the spirit of direct mechanisms (e.g., by multi-stage mechanisms, such as ascending auctions, which implement partial revelation of agent preferences in a series of steps), idiosyncratic constraints on the design problem generally present a more difficult hurdle to overcome.

In this work, we introduce an approach to the design of general mechanisms (direct or indirect) given arbitrary designer objectives and arbitrary constraints on the design space, which we allow to be continuous. We assume that mechanisms induce games of incomplete information in which agents have infinite sets of strategies and types. As in most mechanism design literature, we assume that the designer knows the set of all possible agent types and their distribution, but not the actual type realizations. Our main support for the usefulness of our framework comes from applying it to several problems in auction design which constrain the allocation and/or transfer functions to a particular functional form. In practice, of course, we cannot possibly tackle an arbitrarily complex design space. Our simplification comes from assuming that the designer seeks to find the best setting for particular design parameters. In other words, we allow the designer to search for a mechanism in some subset of an $n$-dimensional Euclidean space, rather than in an arbitrary function space, as would be required in a completely general setting. Furthermore, we believe that many practical design problems involve search for the optimal or nearly optimal setting of parameters within an existing infrustructure. For example, it is much more likely that policy-makers will seek an appropriate tax rate to achieve their objective than overhaul the entire tax system.

In the following sections, we present our framework for automated mechanism design and test it out in several application domains. Our results suggest that our approach has much promise: most of the designs that we discover automatically are nearly as good as or better than the best known hand-built designs in the literature.

## 2 Notation

We restrict our attention to *one-shot games of incomplete information*, denoted by $[I, \{A_i\}, \{T_i\}, F(\cdot), \{u_i(a,t)\}]$, where $I$ refers to the set of players and $m = |I|$ is the number of players. $A_i$ is the set of actions available to player $i \in I$, with $A = A_1 \times \cdots \times A_m$ representing the set of joint actions of all players. $T_i$ is the set of types (private information) of player $i$, with $T = T_1 \times \cdots \times T_m$ representing the joint type space, and $F(\cdot)$ is the joint type distribution. We define a strategy of a player $i$ to be a function $s_i : T_i \to \mathbb{R}$, and use $s(t)$ to denote the vector $(s_1(t_1), \ldots, s_m(t_m))$. It is often convenient to refer to a strategy of player $i$ separately from that of the remaining players. To accommodate this, we use $a_{-i}$ to denote the joint action of all players other than $i$. Similarly, $t_{-i}$ designates the joint type of all players other than $i$. We define the payoff function of each player $i$ by $u_i : A \times T \to \mathbb{R}$, where $u_i(a_i, t_i, a_{-i}, t_{-i})$ is the payoff to player $i$ with type $t_i$ for playing strategy $a_i$ when the remaining players with joint types $t_{-i}$ play $a_{-i}$.

## 3 Automated Mechanism Design Framework

### 3.1 General Framework

We can model the strategic interactions between the designer of the mechanism and its participants as a two-stage game [Vorobeychik et al., 2006]. The designer moves first by selecting a value $\theta$ from a set of allowable mechanism settings, $\Theta$. All the participant agents observe the mechanism parameter $\theta$ and move simultaneously thereafter.

Since the participants know the mechanism parameter, we define a game between them in the second stage as $\Gamma_\theta = [I, \{A_i\}, \{T_i\}, F(\cdot), \{u_i(a, t, \theta)\}]$. We refer to $\Gamma_\theta$ as the game *induced* by $\theta$. As is common in mechanism design literature, we evaluate mechanisms with respect to a sample Bayes-Nash equilibrium, $s(t, \theta)$.[1] We say that given an outcome of play $r$, the designer's goal is to maximize a welfare function $W(r, t, \theta)$ with respect to the distribution of types. Thus, given that a Bayes-Nash equilibrium, $s(t, \theta)$, is the relevant outcome of play, the designer's problem is to maximize $W(s(\theta), \theta) = E_t[W(s(t, \theta), t, \theta)]$.[2]

Observe that if we *knew* $s(t, \theta)$ as a function of $\theta$, the designer would simply be faced with an optimization problem. This insight is actually a consequence of the application of backwards induction, which would have us find $s(t, \theta)$ first for every $\theta$ and then compute an optimal mechanism with respect to these equilibria. If the design space were small, backwards induction applied to our model would thus yield an algorithm for optimal mechanism design. Indeed, if additionally the games $\Gamma_\theta$ featured small sets of players, strategies, and types, we would say little more about the subject. Our goal, however, is to develop a mechanism design tool for settings in which it is infeasible to obtain a solution of $\Gamma_\theta$ for every $\theta \in \Theta$, either because the space of possible mechanisms is large, or because solving (or approximating solutions to) $\Gamma_\theta$ is computationally daunting. Additionally, we try to avoid making assumptions about the objective function or constraints on the design problem or the agent type distributions. We do restrict the games to two players with piecewise linear utility functions, but allow them to have infinite strategy and type sets.

In short, we propose the following high-level procedure for finding optimal mechanisms:

1. Select a candidate mechanism, $\theta$.

2. Find (approximate) solutions to $\Gamma_\theta$.

3. Evaluate the objective and constraints given solutions to $\Gamma_\theta$.

4. Repeat this procedure for a specified number of steps.

---

[1]Focus on a sample equilibrium is typically justified by allowing the designer to suggest the equilibrium to participants, presuming that no agent will subsequently have an incentive to deviate.

[2]Note the overloading of $W(\cdot)$.

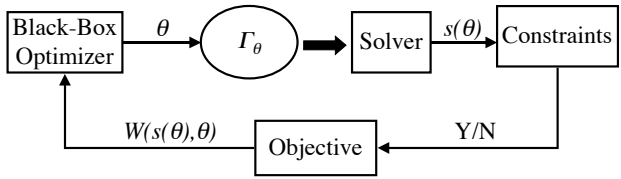

Figure 1: Automated mechanism design procedure based on black-box optimization.

5. Return an approximately optimal design based on the resulting optimization path.

We visually represent this procedure by a diagram in Figure 1.

### 3.2 Designer's Optimization Problem

We begin by treating the designer's problem as black-box optimization, where the black box produces a noisy evaluation of an input design parameter, $\theta$, with respect to the designer's objective, $W(s(\theta), \theta)$, given the game-theoretic predictions of play. Once we frame the problem as a black-box optimization problem, we can draw on a wealth of literature devoted to developing methods to approximate optimal solutions [Spall, 2003]. While we can in principle select a number of these, we have chosen simulated annealing, as it has proved quite effective for a great variety of simulation optimization problems in noisy settings with many local optima.

As an application of black-box optimization, the mechanism design problem in our formulation is just one of many problems that can be addressed with one of a selection of methods. What makes it special is the subproblem of evaluating the objective function for a given mechanism choice, and the particular nature of mechanism design constraints which are evaluated based on Nash equilibrium outcomes and agent types.

### 3.3 Objective Evaluation

As implied by the backwards induction process, we must obtain the solutions (Bayes-Nash equilibria in our case) to the games induced by the design choice, $\theta$, in order to evaluate the objective function. In general, this is simply not possible to do, since Bayes-Nash equilibria may not even exist in an arbitrary game, nor is there a general-purpose tool to find them.

To the best of our knowledge, the only solver for a broad class of infinite games of incomplete information was introduced by Reeves and Wellman [2004] (henceforth, RW). Indeed, RW is a best-response finder, which has successfully been used iteratively to obtain sample Bayes-Nash equilibria for a restricted class of infinite two-player games of incomplete information.

Since the goal of automated mechanism design is to approximate solutions to design problems with arbitrary objectives and constraints and to handle games with arbitrary type distributions, we treat the probability distribution over player types as a black box from which we can sample joint player types. Thus, we use numerical integration (sample mean in our implementation) to evaluate the expectation of the objective with respect to player types, thereby introducing noise into objective evaluation.

### 3.4 Dealing with Constraints

Mechanism design can feature any of the following three classes of constraints: *ex ante* (constraints evaluated with respect to the joint distribution of types), *ex interim* (evaluated separately for each player and type with respect to the joint type distribution of other players), and *ex post* (evaluated for every joint type profile). When the type space is infinite we of course cannot numerically evaluate any expression for every type. We therefore replace these constraints with probabilistic constraints that must hold for "most" types (i.e., a set of types with large probability measure). Intuitively, it is unlikely to matter if a constraint fails for types that occur with probability zero. We conjecture, further, that in most practical design problems, violation of a constraint on a "small" set of types will also be of little consequence, either because the resulting design is easy to fix, or because the other types will likely not have very beneficial deviations even if they account in their decisions for the effect of these unlikely types on the game dynamics. We support this conjecture via a series of applications of our framework: in none of these did our constraint relaxation lead the designer much astray.

Even when we weaken constraints based on agent type sets to their probabilistic equivalents, we still need a way to verify that such constraints hold by sampling from the type distribution. Since we can take only a finite number of samples, we will in fact verify a probabilistic constraint only at some level of confidence. The question we want to ask, then, is how many samples do we need in order to say with probability at least $1 - \alpha$ that the probability of seeing a type profile for which the constraint is violated is no more than $p$? That is the subject of the following theorem.

**Theorem 1.** *Let $B$ denote a set on which a probabilistic constraint is violated, and suppose that we have a uniform prior over the interval $[0, 1]$ on the probability measure of $B$. Then, we need at least $\frac{\log \alpha}{\log(1-p)} - 1$ samples to verify with probability at least $1 - \alpha$ that the* *measure of B is at most p.*

The proofs of this and other results can be found in the appendix of the full version of this paper.

We next describe three specific constraints employed in our applications.

**Equilibrium Convergence Constraint** Given that our game solutions are produced by a heuristic (iterative best-response) algorithm, they are not inherently guaranteed to represent equilibria of the candidate mechanism. We can instead enforce this property through an explicit constraint. The problem that we cannot in practice evaluate this constraint for every joint type profile is resolved by making it probabilistic, as described above. Thus, we define a $(1-p)$-*strong equilibrium convergence constraint*:

**Definition 1.** *Let $s(t)$ be the last strategy profile produced in a sequence of solver iterations, and let $s'(t)$ immediately precede $s(t)$ in this sequence. Then the $(1-p)$-strong equilibrium convergence constraint is satisfied if for a set of type profiles $t$ with probability measure no less than $1-p$, $|s(t) - s'(t)| < \delta$ for some a priori fixed tolerance level $\delta$.*

**Ex Interim Individual Rationality** Ex-Interim-IR specifies that for every agent and type, that agent's expected utility conditional on its type is greater than its opportunity cost of participating in the mechanism. Again, in the automated mechanism design framework, we must change this to a probabilistic constraint as described above.

**Definition 2.** *The $(1-p)$-strong Ex-Interim-IR constraint is satisfied when for every agent $i \in I$, and for a set of types $t_i \in T_i$ with probability measure no less than $1-p$, $E_{t_{-i}} u_i(t, s(t)|t_i) \geq C_i(t_i) - \delta$, where $C_i(t_i)$ is the opportunity cost of agent $i$ with type $t_i$ of participating in the mechanism, and $\delta$ is some a priori fixed tolerance level.*

Commonly in the mechanism design literature the opportunity cost of participation, $C_i(t_i)$, is assumed to be zero but this assumption may not hold, for example, in an auction where not participating would be a give-away to competitors and entail negative utility.

**Minimum Revenue Constraint** The final constraint that we consider ensures that the designer will obtain some minimal amount of revenue (or bound its loss) in attaining a non-revenue-related objective.

**Definition 3.** *The minimum revenue constraint is satisfied if $E_t k(s(t), t) \geq C$, where $k(s(t), t)$ is the total payment made to the designer by agents with joint strategy $s(t)$ and joint type profile $t$, and $C$ is the lower bound on revenue.*

## 4 Extended Example: Shared-Good Auction (SGA)

### 4.1 Setup

Consider the problem of two people trying to decide between two options. Unless both players prefer the same option, no standard voting mechanism (with either straight votes or a ranking of the alternatives) can help with this problem. Instead we propose a simple auction: each player submits a bid and the player with the higher bid wins, paying some function of the bids to the loser in compensation.

We define a space of mechanisms for this problem that are all budget balanced, individually rational, and (assuming monotone strategies) socially efficient. We then search the mechanism space for games that satisfy additional properties. The following is a payoff function defining a space of games parametrized by the function $f$.

$$u(t, a, t', a') = \begin{cases} t - f(a, a') & \text{if } a > a' \\ 0.5t & \text{if } a = a' \\ f(a', a) & \text{if } a < a', \end{cases} \quad (1)$$

where $u()$ gives the utility for an agent who has a value $t$ for winning and chooses to bid $a$ against an agent who has value $t'$ and bids $a'$. The $t$s are the agents' types and the $a$s their actions. Finally, $f()$ is some function of the two bids.[3] In the tie-breaking case (which occurs with probability zero for many classes of strategies) the payoff is the average of the two other cases, i.e., the winner is chosen by a coin flip.

We now consider a restriction of the class of mechanisms defined above.

**Definition 4.** $SGA(h, k)$ *is the mechanism defined by Equation 1 with $f(a, a') = ha + ka'$, $h, k \in [0, 1]$.*

For example, in $SGA(1/2, 0)$ the winner pays half its own bid to the loser. More generally, $h$ and $k$ will be the relative proportions of winner's and loser's bids that will be transferred from the winner to the loser. We now give Bayes-Nash equilibria for such games when types are uniform.

**Theorem 2.** *For $h, k \geq 0$ and types $U[A, B]$ with $B \geq A + 1$ the following is a symmetric Bayes-Nash equilibrium of $SGA(h, k)$:*

$$s(t) = \frac{t}{3(h+k)} + \frac{hA + kB}{6(h+k)^2}$$

For the following discussion, we need to define the notion of truthfulness, or Bayes-Nash incentive compatibility.

---
[3] Reeves [2005] considered the case $f(a, a') = a/2$.

**Definition 5** (BNIC). *A mechanism is Bayes-Nash incentive compatible (truthful) if bidding $s(t) = t$ constitutes a Bayes-Nash equilibrium of the game induced by the mechanism.*

The Revelation Principle [Mas-Colell et al., 1995] guarantees that for any mechanism, there exists a BNIC mechanism that is equivalent in terms of how it maps preferences to outcomes.

We can now characterize the truthful mechanisms in this space. According to Theorem 2, SGA$(1/3, 0)$ is truthful for $U[0, B]$ types. We now show that this is the *only* truthful design in this design space.

**Theorem 3.** *With $U[0, B]$ types ($B > 0$), SGA$(h, k)$ is BNIC if and only if $h = 1/3$ and $k = 0$.*

Below, we show concrete examples of the failure of the revelation principle for several sensible designer objectives.

From here on we restrict ourselves to the case of $U[0, 1]$ types. Since SGA$(1/3, 0)$ is the only truthful mechanism in our design space, we can directly compare the objective value obtained from this mechanism and the best indirect mechanism in the sections that follow.

### 4.2 Automated Design Problems

**Minimize Difference in Expected Utility** First, we consider as our objective *fairness*, or negative differences between the expected utility of winner and loser. Alternatively, our goal is to minimize

$$|E_{t,t'}[u(t, s(t), t', s(t'), k, h \mid a > a') - u(t, s(t), t', s(t'), k, h \mid a < a')]|. \quad (2)$$

We first use the equilibrium bid derived above to analytically characterize optimal mechanisms.

**Theorem 4.** *The objective value in (2) for SGA$(h, k)$ is $(2h + k)/9(h + k)$. Furthermore, SGA$(0, k)$, for any $k > 0$, minimizes the objective, and the optimum is $1/9$.*

By comparison, the objective value for the truthful mechanism, SGA$(1/3, 0)$, is $2/9$, twice as high as the minimum produced by an untruthful mechanism. Thus, the revelation principle does not hold for this objective function in our design space. We can use Theorem 4 to find that the objective value for SGA$(1/2, 0)$, the mechanism described by Reeves [2005], is $2/9$.

Now, to test our framework, we imagine we do not know about the above analytic derivations (including the derivation of the Bayes-Nash equilibrium) and run the automated mechanism design procedure in blackbox mode. Table 1 presents results when we start the search at random values of $h$ and $k$ (taking the best outcome from 5 random restarts), and at the starting values of $h = 0.5$ and $k = 0$. Since the objective function turns out to be fairly simple, it is not surprising that we obtain the optimal mechanism for specific and random starting points (indeed, the optimal design was produced from every random starting point we generated).

| Parameters | Initial Design | Final Design |
|---|---|---|
| $h$ | 0.5 | 0 |
| $k$ | 0 | 1 |
| objective | 2/9 | 1/9 |
| $h$ | random | 0 |
| $k$ | random | 1 |
| objective | N/A | 1/9 |

Table 1: Design that approximately maximizes fairness (minimizes difference in expected utility between utility of winner and loser) when the optimization search starts at a fixed starting point, and the best mechanism from five random restarts.

**Minimize Expected (Ex-Ante) Difference in Utility** Here we modify the objective function slightly as compared to the previous section, and instead aim to minimize the expected ex ante difference in utility:

$$E|u(t, s(t), t', s(t'), k, h|a > a') - u(t, s(t), t', s(t'), k, h|a < a')|. \quad (3)$$

While the only difference from the previous section is the placement of the absolute value sign inside the expectation, this difference complicates the analytic derivation of the optimal design considerably. Therefore, we do not present the actual optimum design values.

| Parameters | Initial Design | Final Design |
|---|---|---|
| $h$ | 0.5 | 0.49 |
| $k$ | 0 | 1 |
| objective | 0.22 | 0.176 |
| $h$ | random | 0.29 |
| $k$ | random | 0.83 |
| objective | N/A | 0.176 |

Table 2: Design that approximately minimizes expected ex ante difference between utility of winner and loser when the optimization search starts at a random and a fixed starting points.

The results of application of our AMD framework are presented in Table 2. While the objective function in this example appears somewhat complex, it turns out (as we discovered through additional exploration) that there are many mechanisms that yield nearly optimal objective values.[4] Thus, both random restarts as well as a fixed starting point produced essentially the same near-optima. By comparison, the truthful design yields the objective value of about 0.22, which is considerably worse.

**Maximize Expected Utility of the Winner**
Yet another objective in the shared-good-auction domain is to maximize the expected utility of the winner. Formally, the designer is maximizing $E[u(t, s(t), t', s(t'), k, h \mid a > a')]$.

We first analytically derive the characterization of optimal mechanisms.

**Theorem 5.** *The problem is equivalent to finding $(h, k)$ that maximize $4/9 - k/[18(h+k)]$. Thus, $k = 0$ and $h > 0$ maximize the objective, and the optimum is $4/9$.*

| Parameters | Initial Design | Final Design |
|---|---|---|
| $h$ | 0.5 | 0.21 |
| $k$ | 0 | 0 |
| objective | 4/9 | 4/9 |
| $h$ | random | 0.91 |
| $k$ | random | 0.03 |
| objective | N/A | 0.443 |

Table 3: Design that approximately maximizes the winner's expected utility.

Here again our results in Table 3 are optimal or very nearly optimal, unsurprisingly for this relatively simple application.

Of the examples we considered so far, most turned out to be analytic, and only one we could only approach numerically. Nevertheless, even in the analytic cases, the objective function forms were not trivial, particularly from a blind optimization perspective. Furthermore, we must take into account that even the simple cases are somewhat complicated by the presence of noise, and thus we need not arrive at global optima even in the simplest of settings so long as the number of samples is not very large.

Having found success in the simple shared-good auction setting, we now turn our attention to a series of considerably more difficult problems.

## 5 Applications

We present results from several applications of our automated mechanism design framework to specific two-player problems. One of these problems, finding auctions that yield maximum revenue to the designer, has been studied in a seminal paper by Myerson [1981] in a much more general setting than ours. Another, which seeks to find auctions that maximize social welfare, has also been studied more generally. Additionally, in several instances we were able to derive optima analytically. For all of these we have a known benchmark to strive for. Others have no known optimal design.

In all of our applications player types are independently distributed with uniform distribution on the unit interval. Finally, we used 50 samples from the type distribution to verify Ex-Interim-IR. This gives us 0.95 probability that 94% of types lose no more than the opportunity cost plus our specified tolerance which we add to ensure that the presence of noise does not overconstrain the problem. It turns out that every application that we consider produces a mechanism that is individually rational for all types *with respect to the tolerance level that was set*.

### 5.1 Myerson Auctions

The seminal paper by Myerson [1981] presented a theoretical derivation of revenue maximizing auctions in a relatively general setting. Here, our aim is to find a mechanism with a nearly-optimal value of some given objective function, of which revenue is an example.[5] However, we restrict ourselves to a considerably less general setting than did Myerson, constraining our design space to that described by the parameters $q, k_1, k_2, K_1, k_3, k_4, K_2$ in (4).

$$u(t, a, t', a') = \begin{cases} U_1 & \text{if } a > a' \\ 0.5(U_1 + U_2) & \text{if } a = a' \\ U_2 & \text{if } a < a', \end{cases} \quad (4)$$

where $U_1 = qt - k_1 a - k_2 a' - K_1$ and $U_2 = (1-q)t - k_3 a - k_4 a' - K_2$. We further constrain all the design parameters to be in the interval [0,1]. In standard terminology, our design space allows the designer to choose an allocation parameter, $q$, which determines the probability that the winner (i.e., agent with the winning bid) gets the good, and transfers, which we constrain to be linear in agents' bids.

While our automated mechanism design framework assures us that *p*-strong individual rationality will hold with the desired confidence, we can actually verify it by hand in this application. Furthermore, we can adjust the mechanism to account for lapses in individual rationality guarantees for subsets of agent types by giving to each agent the amount of the expected loss of the least fortunate type.[6] Similarly, if we do find a mechanism that is Ex-Interim-IR, we may still have an opportunity to increase expected revenue as long as the minimum expected gain of any type is strictly greater than zero.

---

[4] Particularly, we carried out a far more intensive exploration of the search space given the analytic expression for the Bayes-Nash equilibrium to ascertain that the values reported are close to actual optima. Indeed, we failed to improve on these.

[5] Conitzer and Sandholm [2003a] also tackled Myerson's problem, but assumed finite type and strategy spaces of agents, as well as a finite design space.

**Maximize Revenue** Here we are interested in finding approximately revenue-maximizing designs in our constrained design space. First, we derive the following based on Myerson's feasibility constraints:

**Theorem 6.** *Optimal incentive compatible mechanism in our setting yields the revenue of 1/3, which can be achieved by selecting $q = 1$, $k_1 \in [0, 0.5]$, and $k_2 \in [0, 1]$, respecting the constraint that $k_1 + 0.5k_2 = 0.5$.*

In addition to performing five restarts from random starting points, we repeated the simulated annealing procedure starting with the best design produced via the random restarts. **This procedure yielded an Ex-Interim-IR design with expected revenue of approximately 0.3.** We used the RW solver to find a symmetric equilibrium of this design, under which the bids are $s(t) = 0.72t - 0.73$.[7] We have already shown that the best known design, which is also the optimal incentive compatible mechanism in this setting, yields a revenue of 1/3 to the designer. Thus, our AMD framework produced a design near to the best known. It is an open question what the actual global optimum is.

**Maximize Welfare** It is well known that the Vickrey auction is welfare-optimal. Thus, we know that the welfare optimum is attainable in our design space. Before proceeding with search, however, we must make one observation. While we are interested in welfare, it would be inadvisable in general to completely ignore the designer's revenue, since the designer is unlikely to be persuaded to run a mechanism at a disproportionate loss. To illustrate, take the same Vickrey auction, but afford each agent one billion dollars for participating. This mechanism is still welfare-optimal, but seems a senseless waste if optimality could be achieved without such spending (and, indeed, at some profit to the auctioneer). To remedy this problem, we use a minimum revenue constraint, ensuring that no mechanism that is too costly will be selected as optimal.

---

[6]Observe that such constant transfers will not affect agent incentives.

[7]This equilibrium is approximate in the sense that we rounded the parameters to the nearest hundredth.

First, we present a general result that characterizes welfare-optimal mechanisms in our setting.

**Theorem 7.** *Welfare is maximized if either the equilibrium bid function is strictly increasing and $q = 1$ or the equilibrium bid function is strictly decreasing and $q = 0$. Furthermore, the maximum expected welfare in our design space is 2/3.*

Thus, for example, both first- and second-price sealed bid auctions are welfare-optimizing (as is well known).

**The result of our search for optimal design is an Ex-Interim-IR mechanism which allocated the object to the highest bidder. This mechanism yields expected revenue of approximately 0.2 to the designer.** We verified using the RW solver that the bid function $s(t) = 0.645t - 0.44$ is an equilibrium given this design. Since it is strictly increasing in $t$, we can conclude based on Theorem 7 that **this design is welfare-optimal.**

### 5.2 Vicious Auctions

In this section we study a design problem motivated by the Vicious Vickrey auction [Brandt and Weiß, 2001]. The essence of this auction is that while it is designed exactly like a regular Vickrey auction, the players get disutility from the utility of the other player, which is a function of parameter $l$, with the regular Vickrey auction the special case of $l = 0$.

We generalize the Vicious Vickrey auction design using the same parameters as in the previous section such that the Vicious Vickrey auction is a special case with $q = k_2 = 1$ and $k_1 = k_2 = k_3 = k_4 = K_1 = K_2 = 0$, and the utility function of agents presented in the previous section can be recovered when $l = 0$. We assume in this construction that payments, which will be the same (as functions of players' bids and design parameters) as in the Myerson auction setting, have a particular effect on players' utility parametrized by $l$. Hence, the utility function in (5).

$$u(t, a, t', a') = \begin{cases} U_1 & \text{if } a > a' \\ 0.5(U_1 + U_2) & \text{if } a = a' \\ U_2 & \text{if } a < a' \end{cases} \quad (5)$$

where $U_1 = q(1-l)t - (k_1(q(1-l) + (1-q)) - (1-q)l)a - ((1-q)l)t' - k_2(q(1-l) + (1-q))a' - K_1$ and $U_2 = (1-q)(1-l)t - (k_3((1-q)(1-l) + q) - ql)a - qlt' - k_4((1-q)(1-l) + q)a' - K_2$. In all the results below, we fix $l = 2/7$. Reeves [2005] reports an equilibrium for Vicious Vickrey with this value of $l$ to be $s(t) = (7/9)t + 2/9$. Thus, we can see that we are no longer assured incentive compatibility even in the second-price auction case. In general, it is unclear whether there exist incentive compatible mechanisms in this design space, particularly because we constrain all our parameters to lie in the interval $[0, 1]$.

In the applications below, we redefined the individual rationality constraint in terms of an agent's opportunity cost of participation in the auction.

**Maximize Revenue** Our first objective is to (nearly) maximize revenue in this domain. *Our AMD framework achieves an Ex-Interim-IR mechanism with expected revenue of approximately 0.44. By comparison, a Vicious Vickrey auction achieves revenue of 0.48, but is not Ex-Interim-IR.*

**Maximize Welfare** We now tackle the objective of maximizing welfare using our AMD framework. *The result is a mechanism with expected welfare of approximately 0.54, which is not Ex-Interim-IR. However, the designer can pay each agent 0.065, thereby making the design individually rational, and still maintain positive revenue.*

**Maximize Weighted Sum of Revenue and Welfare** In this section, we present results of AMD with the goal of maximizing the weighted sum of revenue and welfare. For simplicity (and having no reason for doing otherwise), we set weights to be equal. *Our framework found a mechanism which is welfare-optimal and yields revenue of 0.52 after an adjustment which makes it Ex-Interim-IR.* Interestingly, we were much more successful in both revenue and welfare objectives by eliminating the hard minimum revenue constraint and instead making it a part of the objective. Indeed, we found here the best mechanism so far for *both* objectives we considered, suggesting that there is also synergy between the two objectives.

## 6 Conclusion

We presented a framework for automated design of general mechanisms (direct or indirect) using the Bayes-Nash equilibrium solver for infinite games developed by Reeves and Wellman [2004]. Results from applying this framework to several design domains demonstrate the value of our approach for practical mechanism design. The mechanisms that we found were typically either close to the best known mechanisms, or better.

While in principle it is not at all surprising that we can find mechanisms by searching the design space— as long as we have an equilibrium finding tool—it was not at all clear that any such system will have practical merit. We presented evidence that indirect mechanism design in a constrained space can indeed be effectively automated on somewhat realistic design problems that yield very large games of incomplete information. Undoubtedly, real design problems are vastly more complicated than any that we considered (or any that can be considered theoretically). In such cases, we believe that our approach could offer considerable benefit if used in conjunction with other techniques, either to provide a starting point for design, or to tune a mechanism produced via theoretical analysis and computational experiments.